\documentclass[aps,prb,nobibnotes,twocolumn,superscriptaddress,bibliography]{revtex4-2}

\usepackage{amsfonts}
\usepackage{mathrsfs}
\usepackage{amsmath}
\usepackage{color}
\usepackage{graphicx}
\usepackage{bm}
\usepackage{amssymb}
\usepackage{xspace}
\usepackage{epstopdf}
\usepackage{dcolumn}
\usepackage{longtable}
\usepackage{multirow}
\usepackage{float}
\usepackage{comment}

\providecommand{\tabularnewline}{\\}

\usepackage[colorlinks=true, letterpaper=true, pdfstartview=FitV, linkcolor=blue, citecolor=blue, urlcolor=blue]{hyperref}

\makeatother

\begin{document}
\title{Four-band tight-binding model of TiSiCO-family monolayers}
\author{Chaoxi Cui}
\affiliation{Centre for Quantum Physics, Key Laboratory of Advanced Optoelectronic
Quantum Architecture and Measurement (MOE), School of Physics, Beijing
Institute of Technology, Beijing 100081, China}
\affiliation{Beijing Key Lab of Nanophotonics \& Ultrafine Optoelectronic Systems,
School of Physics, Beijing Institute of Technology, Beijing 100081,
China}
\author{Yilin Han}
\affiliation{Centre for Quantum Physics, Key Laboratory of Advanced Optoelectronic
Quantum Architecture and Measurement (MOE), School of Physics, Beijing
Institute of Technology, Beijing 100081, China}
\affiliation{Beijing Key Lab of Nanophotonics \& Ultrafine Optoelectronic Systems, School of Physics, Beijing Institute of Technology, Beijing 100081,
China}
\author{Ting-Ting Zhang}
\affiliation{Beijing National Laboratory for Condensed Matter Physics, Institute of Physics, Chinese Academy of Sciences, Beijing 100190, China }
\author{Zhi-Ming Yu}
\email{zhiming\_yu@bit.edu.cn}
\affiliation{Centre for Quantum Physics, Key Laboratory of Advanced Optoelectronic Quantum Architecture and Measurement (MOE), School of Physics, Beijing
Institute of Technology, Beijing 100081, China}
\affiliation{Beijing Key Lab of Nanophotonics \& Ultrafine Optoelectronic Systems, School of Physics, Beijing Institute of Technology, Beijing 100081, China}

\author{Yugui Yao}
\affiliation{Centre for Quantum Physics, Key Laboratory of Advanced Optoelectronic Quantum Architecture and Measurement (MOE), School of Physics, Beijing
Institute of Technology, Beijing 100081, China}
\affiliation{Beijing Key Lab of Nanophotonics \& Ultrafine Optoelectronic Systems, School of Physics, Beijing Institute of Technology, Beijing 100081, China}

\begin{abstract}
The TiSiCO-family monolayers have recently been attracting significant attention due to their unique valley-layer coupling (VLC).
In this work, we present a minimal, four-band tight-binding (TB) model to capture  the low-energy physics of the TiSiCO-family monolayers $X_{2}Y$CO$_{2}$
($X=$ Ti, Zr, Hf; $Y=$ Si, Ge) with strong VLC.
These monolayers comprise two $X$ atom layers separated by approximately $4$ \AA ~in the out-of-plane direction.
Around each valley ($X$ or $X'$), the conduction  and valence bands are mainly dominated by the $A_{1}\{d_{z^{2}(x^{2}-y^{2})}\}$ and $B_{2}\{d_{yz}\}$ orbitals of the top $X$ atoms,and  the $A_{1}\{d_{z^{2}(x^{2}-y^{2})}\}$ and $B_{1}\{d_{xz}\}$ orbitals of the bottom $X$ atoms.
Using these four states as a basis, we construct a symmetry-allowed TB model.
Through parameter fitting from first-principles calculations, the four-band TB model not only reproduces the electronic band structure, but also captures the strong VLC, high-order topology, and valley-contrasting linear dichroism of the monolayers.
Furthermore, the TB model reveals that these monolayers may exhibit various intriguing topological phases under electric fields and biaxial strains.
Hence, the TB model established here  can serve as the starting point for future research exploring the physics related to VLC and the $X_{2}Y$CO$_{2}$ monolayers.
\end{abstract}
\maketitle

\section{Introduction }

Valleytronics materials, which are characterized by the presence of multiple symmetry-connected energy extremal points in the low-energy
bands, have been a focus of research in condensed matter physics \cite{2007_Beenakker_NP,2006_Valleytronic_earliest2D,RN816,2007_XiaoDi,2012_XiaoDi,2006_Valleytronic_earliest2D,2007_XiaoDi,2008_YaoWang,2012_XiaoDi,cai2013magnetic,2015_PanHui}.
The concept of valleytronics works in both  three dimensions  and two dimensions. However, due to the flexibility and controllability in two dimensions, it is the discovery of two-dimensional (2D)  valleytronics materials
like graphene and transitional metal dichalcogenides (TMDs) that leads to the rapid growth in the field of valleytronics~\cite{Valleytronics2016review,Valleytronics2018review}. The 2D valleytronics
materials are particularly attractive for both fundamental studies
and the development of application devices~\cite{2012_ZengweiZhu,MagneticControl_Experiment,ValleyZeem_Experiment,2014_XuXiaodong,gunlycke2011graphene,pan2015perfect,pan2015valley,jiang2013generation,grujic2014spin,ju2015topological,sui2015gate,tong2016concepts,nguyen2016valley,akhmerov2008theory,cresti2008valley,san2009pseudospin,qiao2011electronic,qiao2014current,li2018valley,garcia2008fully,settnes2016graphene,cheng2018manipulation}.

Recently, the TiSiCO-family monolayers $X_{2}Y$CO$_{2}$  ($X=$Ti, Zr, Hf; $Y$=Si, Ge)  have been proposed as a novel class of 2D valleytronics materials~\cite{VLC_Yu}.
Similar to the graphene and TMDs,  both the conduction and valence bands of these monolayers exhibit two valleys located at two high-symmetry
points of the Brillouin zone (BZ), namely, $X$ and $X'\ (Y)$ points. However, the two valleys in monolayer  $X_{2}Y$CO$_{2}$ are time-reversal ${\cal{T}}$ invariant points and are connected by the spatial operators $S_{4z}$ and $C_{2,110}$, which is completely different from that in the graphene and TMDs~\cite{zhang2011spontaneous,YaoWang2009graphene,Guibin_MoS2,2012_XiaoDi,guibinreview,GrapheneRMP}. As a result, the valley polarization in monolayer  $X_{2}Y$CO$_{2}$ can be realized by the methods that do not break ${\cal{T}}$ symmetry.
Particularly, due to the strong VLC--the conduction or valence electrons in different valleys have strong but opposite layer polarization, the  gate electric field is an intuitive and efficient way to generate valley polarization in monolayer  $X_{2}Y$CO$_{2}$.
This electric control of valley polarization  is highly desirable for the applications.
In addition to static control, dynamical generation of valley polarization also  can  be realized in these monolayers, as they exhibit valley-contrasting linear
dichroism \cite{VLC_Yu}.
Furthermore, it has been predicted that the monolayer  $X_{2}Y$CO$_{2}$ is not a normal semiconductor but a second-order topological insulator~[cite han].
Therefore, monolayer $X_{2}Y$CO$_{2}$  will be of broad interest to multiple fields, including  valleytronics, 2D materials, optoelectronics, and higher-order topology.

In our previous work, an effective two-band $\boldsymbol{k\cdot p}$ model was developed based on invariant theory \cite{VLC_Yu}, where the spin-orbit coupling (SOC) effect is not included due to negligible SOC in the low-energy bands of  the monolayer  $X_{2}Y$CO$_{2}$. The effective model clearly demonstrates the coupling between valley and layer degrees of freedom, and can be used to describe the optical properties of the monolayer  $X_{2}Y$CO$_{2}$. However, it is insufficient to capture the higher-order topology of systems and the physics away from the two valleys.

In this work, we present a minimal lattice  model for the monolayer  $X_{2}Y$CO$_{2}$ without SOC effect. The TB model is constructed by the $d$ orbitals of $X$ atoms, i.e. the $A_{1}\{d_{z^{2}(x^{2}-y^{2})}\}$ and $B_{2}\{d_{yz}\}$ orbitals of the top $X$ atoms, and the $A_{1}\{d_{z^{2}(x^{2}-y^{2})}\}$
and $B_{1}\{d_{xz}\}$ orbitals of the bottom $X$ atoms. This effective model contains four bands: two valence bands and two conduction bands. All parameters in the model are obtained by fitting the electronic bands from the first-principles calculations.
We demonstrate that our four-band TB model can effectively describe the low-energy physics of the monolayer $X_{2}Y$CO$_{2}$, including strong VLC, optical properties, and higher-order topology. Furthermore, the TB model suggests that the monolayer $X_{2}Y$CO$_{2}$ may undergo multiple phase transitions under external fields.

This paper is organized as follows. In Sec. \ref{sec:II}, we introduce
the processes that lead to our four-band lattice model. In Sec. \ref{sec:III},
the optical and topological properties of the effective lattice model
are studied. We investigate possible phase transitions of the model
under external field in Sec. \ref{sec:IV}. Conclusions are given
in Sec. \ref{sec:V}.

\begin{figure}
	\includegraphics[width=8.8cm]{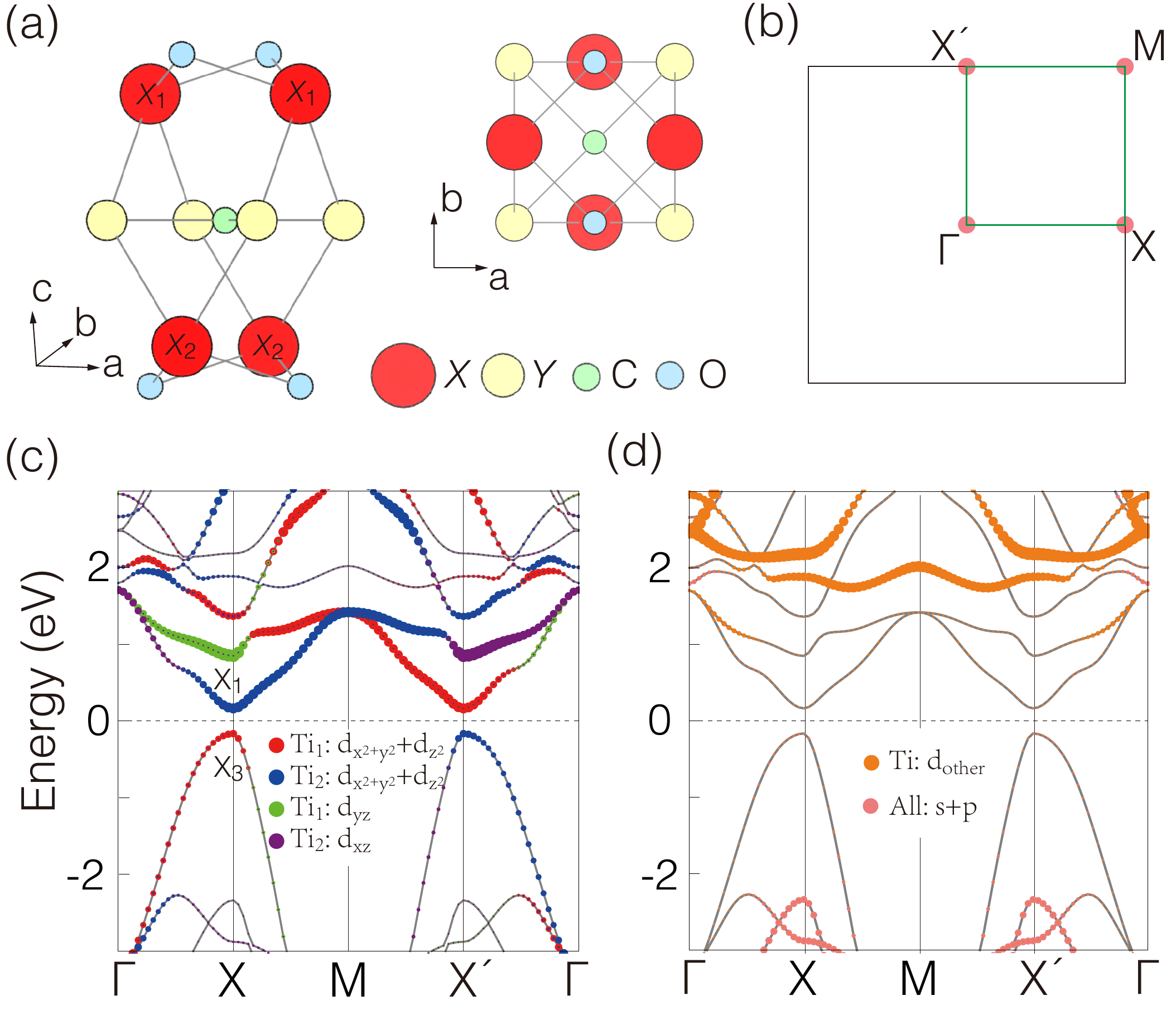}
	\caption{(a) Side and top views of monolayer $X_{2}Y$CO$_{2}$. (b) denotes the BZ of the monolayer.  (c-d) show a representative example of orbital-projected band structures for the monolayer $X_{2}Y$CO$_{2}$ calculated from first-principle calculations. Here, $X$ and $Y$ represent Ti and Si, respectively. (c) presents the orbitals included in the TB model, while (d) displays the other orbitals that are not incorporated in the model.}
	\label{Fig.1}
\end{figure}

\section{The minimal lattice model \label{sec:II}}

The monolayer  $X_{2}Y$CO$_{2}$ belongs to layer group (LG) NO. 59 or space group (SG) No. 115 with $D_{2d}$ point-group symmetry.
The crystalline structure of the monolayer  $X_{2}Y$CO$_{2}$ is shown in Fig. \ref{Fig.1}(a) and the electronic band of the monolayer Ti$_2$SiCO$_{2}$ is plotted in Fig. \ref{Fig.1}(c-d). Figure \ref{Fig.1}(b) shows the Brillouin zone (BZ) with the high-symmetry points being labeled. Here, the position of the $X$ point is $(\pi,0)$ and that of $X'$ point is $(0,\pi)$.
Notice that there exists  an alternative notation \cite{website} where  the positions of $X$ and $X'$ points are interchanged, which is adopted in our previous work \cite{VLC_Yu}.
The $X$, $Y$, C, and O atoms are located at $2g$, $1b$, $1a$ and $2g$ Wyckoff positions, respectively.

{\global\long\def\arraystretch{1.4}%
\begin{table}
\caption{The first two columns show the compatibility relaxation between $d$-orbitals and the irreducible representations (IRRs) of $C_{2v}$ point group. The last column presents the band representation (BR) of SG 115 from $2g$ Wyckoff position, for which the site symmetry is $C_{2v}$.\label{tab:1}}
\begin{ruledtabular} %
\begin{tabular}{lllllllll}
\multirow{2}{*}{$d$-orbital} & \multirow{2}{*}{IRRs} & \multicolumn{4}{c}{Induced BRs ($R\uparrow G$)}\tabularnewline
 &  & $R\uparrow G$ & $\Gamma$ & $X$ & $M$\tabularnewline
\hline
$d_{z^{2}}$, $d_{x^{2}-y^{2}}$ & $A_{1}$ &  $A_{1}\uparrow G$  & $\Gamma_{1}\oplus\Gamma_{2}$ & $X_{1}\oplus X_{3}$ & $M_{5}$\tabularnewline
$d_{xy}$ & $A_{2}$ & $A_{2}\uparrow G$  &  $\Gamma_{3}\oplus\Gamma_{4}$ & $X_{2}\oplus X_{4}$ & $M_{5}$\tabularnewline
$d_{xz}$ & $B_{1}$ & $B_{1}\uparrow G$  &  $\Gamma_{5}$ & $X_{2}\oplus X_{3}$ & $M_{3}\oplus M_{4}$\tabularnewline
$d_{yz}$ & $B_{2}$ & $A_{2}\uparrow G$  &  $\Gamma_{5}$ & $X_{1}\oplus X_{4}$ & $M_{1}\oplus M_{2}$\tabularnewline
\end{tabular}\end{ruledtabular}
\end{table}
}

Around Fermi level, the electronic bands of these monolayers mainly consist of certain $d$ orbitals of the two $X$ atoms, while  the contribution of other orbits, i.e., other $d$ orbit of $X$ atoms, and $s$ and $p$ orbit of all other atoms is negligible, as shown in  Fig. \ref{Fig.1}(c-d) and appendix (Fig. \ref{Fig.S1}). Specifically, the low-energy bands at $X$ ($X'$) valley are dominated by the $d_{yz}$ ($d_{xz}$) orbits of the top (bottom) $X$ atoms, and the $d_{z^{2}}$ and $d_{x^{2}-y^{2}}$ orbits of the both top and bottom $X$ atoms [see Fig. \ref{Fig.1}(c)].
From these band analysis, one knows that the valley states including both conduction and valence valley states have strong layer polarization, and the layers polarization for the $X$ and $X'$ valleys are opposite, leading to strong VLC effect. The VLC effect is protected by the  $S_{4z}$ and $C_{2,110}$ symmetries.
The band representation of the conduction (valence) band edge at $X$ valley is calculated as $X_{1}$ ($X_{3}$) \cite{GuibinMSGrep}, from which  the band representation of the band edges at $X'$ valley can be inferred.

\begin{table*}
	\caption{Fitted parameters of the four-band TB model for monolayer $X_{2}Y$CO$_{2}$ with inter-layer and intra-layer NN hoppings based on first-principle calculations.  ``$t$" and  ``$r$" stand for inter-layer NN and intra-layer NN hoppings, respectively. All parameters are in unit of eV.}
\begin{ruledtabular} %
	\begin{tabular}{ccccccccccc}
		& $\Delta$ & $t_{1,x}$ & $t_{1,y}$ & $t_{2,x}$ & $t_{2,y}$ & $t^{\prime}$ & $r_{1}$ & $r_{2}$ & $r^{\prime}$\tabularnewline
		\hline
		TiSiCO & 0.036 & -0.729 & -0.866 & 0.621 & 1.113 & 0.677 & -0.355 & -0.322 & -0.127\tabularnewline
		TiGeCO & 0.137 & -0.883 & -1.058 & 0.781 & 1.323 & 0.768 & -0.424 & -0.300 & -0.221\tabularnewline
		ZrSiCO & 0.063 & -0.927 & -1.245 & 0.699 & 1.593 & 1.047 & -0.502 & -0.418 & -0.323\tabularnewline
		ZrGeCO & 0.060 & -1.007 & -1.279 & 0.702 & 1.535 & 0.981 & -0.527 & -0.364 & -0.305\tabularnewline
		HfSiCO & 0.021 & -1.062 & -1.440 & 0.799 & 1.891 & 1.166 & -0.644 & -0.510 & -0.381\tabularnewline
		HfGeCO & 0.042 & -1.144 & -1.446 & 0.698 & 1.587 & 1.082 & -0.682 & -0.408 & -0.332\tabularnewline
	\end{tabular}
\end{ruledtabular}
\label{tab.2}
\end{table*}

The site symmetry group of the $X$ atoms (Wyckoff position $2g$ in SG No. 115) is $C_{2v}$. For spinless systems, the $d$ orbitals in $C_{2v}$ point-group symmetry would split into five non-degenerate energy levels: $2A_{1}+A_{2}+B_{1}+B_{2}$, as listed in Table \ref{tab:1}. Since the $d_{z^{2}}$ and $d_{x^{2}-y^{2}}$ orbitals share the same representation $A_{1}$ of the $C_{2v}$ point group,
it is unnecessary to distinguish them in the band analysis. Therefore, for simplification, we only use the $d_{z^{2}}$ and $d_{yz}$ orbitals of the top $X$ atom and the $d_{z^{2}}$ and $d_{xz}$ orbitals of the bottom $X$ atom to construct a four-band model. It can be proved that the four-band model is the minimal one to capture the physics of the monolayer $X_{2}Y$CO$_{2}$. First, since there are two $X$ atoms in a unit cell,  the band number of the lattice model must be even. Second, the band representations of the SG No. 115 (with  ${\cal T}$ symmetry) from $2g$ Wyckoff position can be found in the BCS website \cite{website,TopologicalQuantumChemistry}, and are rewritten in Table \ref{tab:1}. From Table \ref{tab:1}, one observes that a two-band lattice model based on the $d$ orbitals of $X$ atom must be a semimetal, as the two bands will be degenerate at  $\Gamma$ or $M$ points. However, the monolayer $X_{2}Y$CO$_{2}$ is a semiconductor rather than a semimetal. This contradiction indicates that the lattice model of the monolayer $X_{2}Y$CO$_{2}$ should have (at least) four bands.

\begin{figure*}
	\includegraphics[width=14cm]{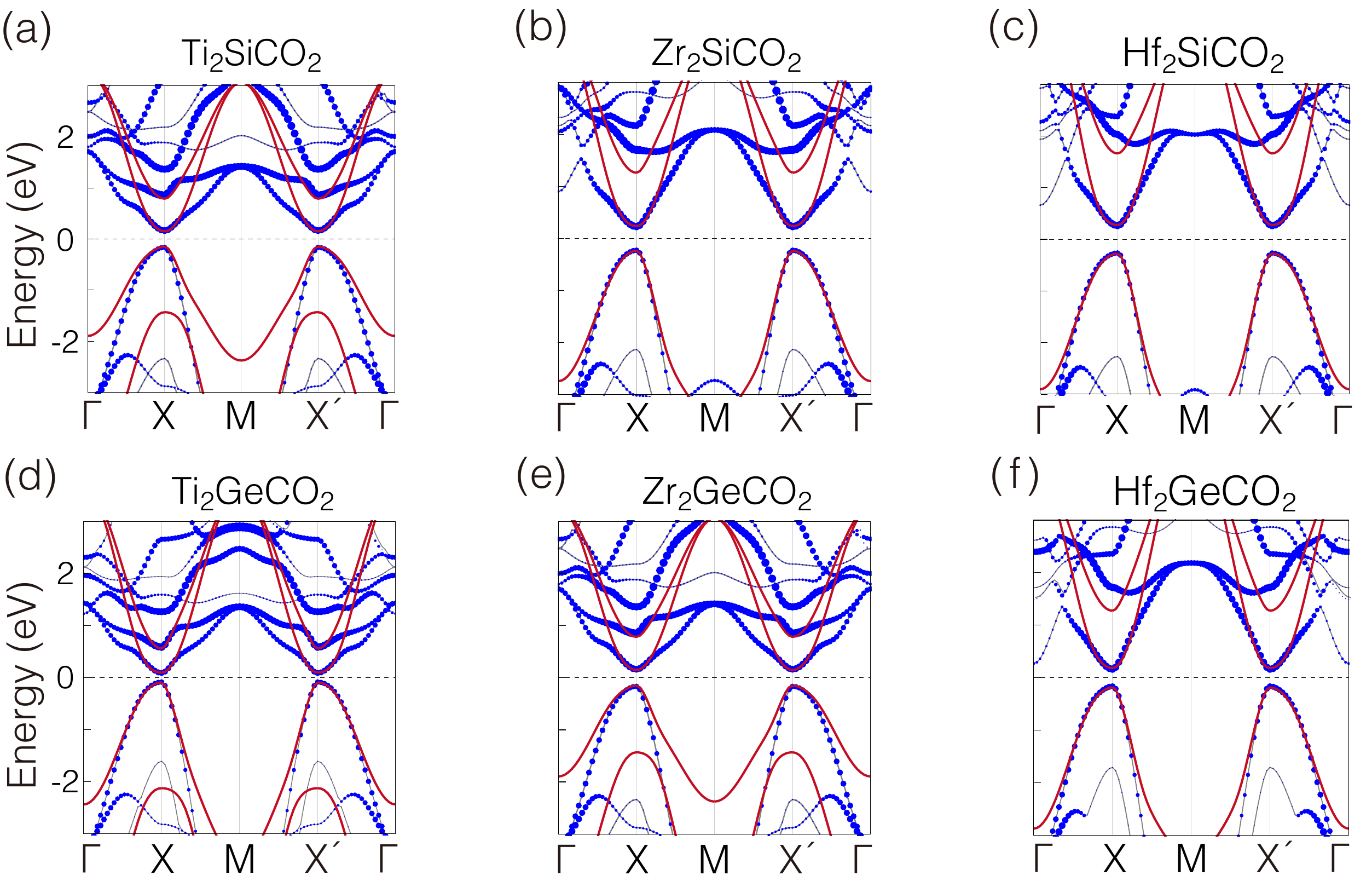}
	\caption{(a-f) The TB band structures (red) for monolayer $X_{2}Y$CO$_{2}$ are compared with the first-principles band structures (gray). The blue dots indicate the band components of orbitals included in the TB model.}
	\label{Fig.2}
\end{figure*}

To construct the lattice  model, we need to determine the matrix representations of the generators of the SG 115. The basis of the TB model here is  $\{d_{z^{2}}^{1},d_{yz}^{1},d_{z^{2}}^{2},d_{xz}^{2}\}$, where the superscript $1(2)$ denotes the top (bottom) $X$ atom, located at $(0,b/2,d/2)$ and $(a/2,0,-d/2)$, respectively. Here, $a=b$ is  the lattice constant and $d$ refers to  the vertical distance  between the two $X$ atoms. The generators of the symmetry operators of the monolayer $X_{2}Y$CO$_{2}$ are $S_{4z}$, $M_{y}$ and $\mathcal{T}$,  and their matrix representations are obtained as: \begin{eqnarray}
S_{4z}=\left[\begin{array}{cccc}
0 & 0 & -1 & 0\\
0 & 0 & 0 & 1\\
-1 & 0 & 0 & 0\\
0 & -1 & 0 & 0
\end{array}\right], & \  & M_{y}=\left[\begin{array}{cccc}
1 & 0 & 0 & 0\\
0 & -1 & 0 & 0\\
0 & 0 & 1 & 0\\
0 & 0 & 0 & 1
\end{array}\right],
\end{eqnarray}
and
\begin{equation}
\mathcal{T}=\left[\begin{array}{cccc}
1 & 0 & 0 & 0\\
0 & 1 & 0 & 0\\
0 & 0 & 1 & 0\\
0 & 0 & 0 & 1
\end{array}\right]\mathcal{K},
\end{equation}
where $\mathcal{K}$ is complex conjugation operator. Then, the symmetry-allowed
TB Hamiltonian of the monolayer $X_{2}Y$CO$_{2}$ is established
as
\begin{eqnarray}
{\cal H} & = & \left(\begin{array}{cc}
H_{\rm top} & H_{\rm inter}\\
H_{\rm inter}^{\dagger} & H_{\rm bottom}
\end{array}\right),\label{eq:ham}
\end{eqnarray}
with $H_{\rm bottom}\left(k_{x},k_{y}\right)=H_{\rm top}\left(k_{y},-k_{x}\right)$,
\begin{eqnarray*}
H_{\rm top}=\Delta \sigma_z+\left(\begin{array}{cc}
	\sum\limits_{\alpha=x,y}t_{1,\alpha}\cos k_{\alpha} & it^{\prime}\sin k_{y}\\
	-it^{\prime}\sin k_{y} & \sum\limits_{\alpha=x,y}t_{2,\alpha}\cos k_{\alpha}
\end{array}\right),
\end{eqnarray*}
and
\begin{eqnarray*}
	H_{\rm inter}=\left(\begin{array}{cc}
		r_{1}\cos\frac{k_{x}}{2}\cos\frac{k_{y}}{2} & ir^{\prime}\cos\frac{k_{x}}{2}\sin\frac{k_{y}}{2}\\
		ir^{\prime}\cos\frac{k_{y}}{2}\sin\frac{k_{x}}{2} & r_{2}\sin\frac{k_{x}}{2}\sin\frac{k_{y}}{2}
	\end{array}\right).
\end{eqnarray*}
Here, $\sigma_z$ is the $z$-component of the Pauli matrix, and all the parameters are real. The Hamiltonian (\ref{eq:ham}) contains the
nearest-neighbor (NN) intra-layer and NN inter-layer hoppings, resulting in $9$ symmetry-allowed real parameters. We employ
the gradient descent method \cite{lemarechal2012cauchy} to determine the parameters by comparing the electronic bands from the first-principles calculations with those from the TB model (\ref{eq:ham})  with appropriate initial parameters. The fitted bands of the monolayer  $X_{2}Y$CO$_{2}$  are illustrated  in Fig. \ref{Fig.2} and the corresponding  parameters are listed in Table \ref{tab.2}.

As studied in Ref. \cite{VLC_Yu}, one of the most intriguing  properties of the monolayer $X_{2}Y$CO$_{2}$ is the strong VLC effect.  To demonstrate that our TB model can capture the VLC effect, we present the layer polarization of the valley states of the TB model in Fig. \ref{Fig.3}. The layer polarization is defined as \cite{VLC_Yu}
\begin{eqnarray}
P_{n}(\boldsymbol{k}) & = & \int_{z>0}\left|\psi_{n\boldsymbol{k}}\right|^{2}d\boldsymbol{r}-\int_{z<0}\left|\psi_{n\boldsymbol{k}}\right|^{2}d\boldsymbol{r},
\end{eqnarray}
with $\psi_{n\boldsymbol{k}}$ representing the eigenstate of $n$-th Bloch band
and $\boldsymbol{k}$ denoting the wave vector. Here, the $z=0$ plane is set on the middle $Y$/C atom layer. The layer polarization $P_{n}(\boldsymbol{k})$
indicates the polarization of $\psi_{n\boldsymbol{k}}$ between the
top ($z>0$) and bottom ($z<0$) layers. From Fig. \ref{Fig.3}, strong valley-contrasted layer polarization can be observed for  both conduction and valence bands, reproducing the VLC effect in the monolayer $X_{2}Y$CO$_{2}$.

\section{Optical and topological properties \label{sec:III}}

In this section, we show that the four-band TB model can describe the optical and topological properties of the monolayer $X_{2}Y$CO$_{2}$. TMDs are known to exhibit valley-contrasting circular dichroism in optical interband absorption~\cite{CircularDichroism,2012_XiaoDi,guibinreview}. However, due to the difference in symmetry, the $X$ ($X'$) valleys in monolayer $X_{2}Y$CO$_{2}$
exclusively couple to $x$-linearly ($y$-linearly) polarized light \cite{VLC_Yu} rather than the circularly polarized light. Consequently, the monolayer $X_{2}Y$CO$_{2}$
features valley-contrasting linear dichroism. The $\boldsymbol{k}$-resolved
linear polarization degree of the optical interband absorption between valence
and conduction bands is characterized by
\begin{eqnarray}
\eta(\boldsymbol{k}) & =\frac{\left|M_{x}\right|^{2}-\left|M_{y}\right|^{2}}{\left|M_{x}\right|^{2}+\left|M_{y}\right|^{2}} &
\end{eqnarray}
where $M_{i}=\langle\psi_{c \mathrm{k}}|\frac{\partial H}{\partial k_i}| \psi_{v \mathrm{k}}\rangle$ is the coupling strength  between valence and conduction band with the optical field linearly polarized in the $i$-th direction. $\eta(\boldsymbol{k})$ indicates the normalized absorption difference between $x$ and $y$-linearly polarized light. The $\eta(\boldsymbol{k})$ calculated from our TB model is shown in Fig. \ref{Fig.4}(b). We find that  $\eta(X)=1$ and $\eta(X')=-1$,  indicating an opposite linear dichroism around the $X$ and $X'$ valleys. This is  consistent with the results in Ref. {\cite{VLC_Yu}}.
Moreover, since the four high-symmetry lines, $\Gamma$-X, $\Gamma$-$X'$, $M$-$X$, and $M$-$X'$ have mirror symmetry ($M_x$ or $M_y$), the electronic states on them  must exclusively  couple to a  linearly polarized light whose polarization direction is either parallel or perpendicular to the mirror.
This property also can be found in our calculations, where  $\eta({\bm k})=\pm 1$ for the four high-symmetry lines [see Fig. \ref{Fig.4}(b)].

\begin{figure}
	\includegraphics[width=8.8cm]{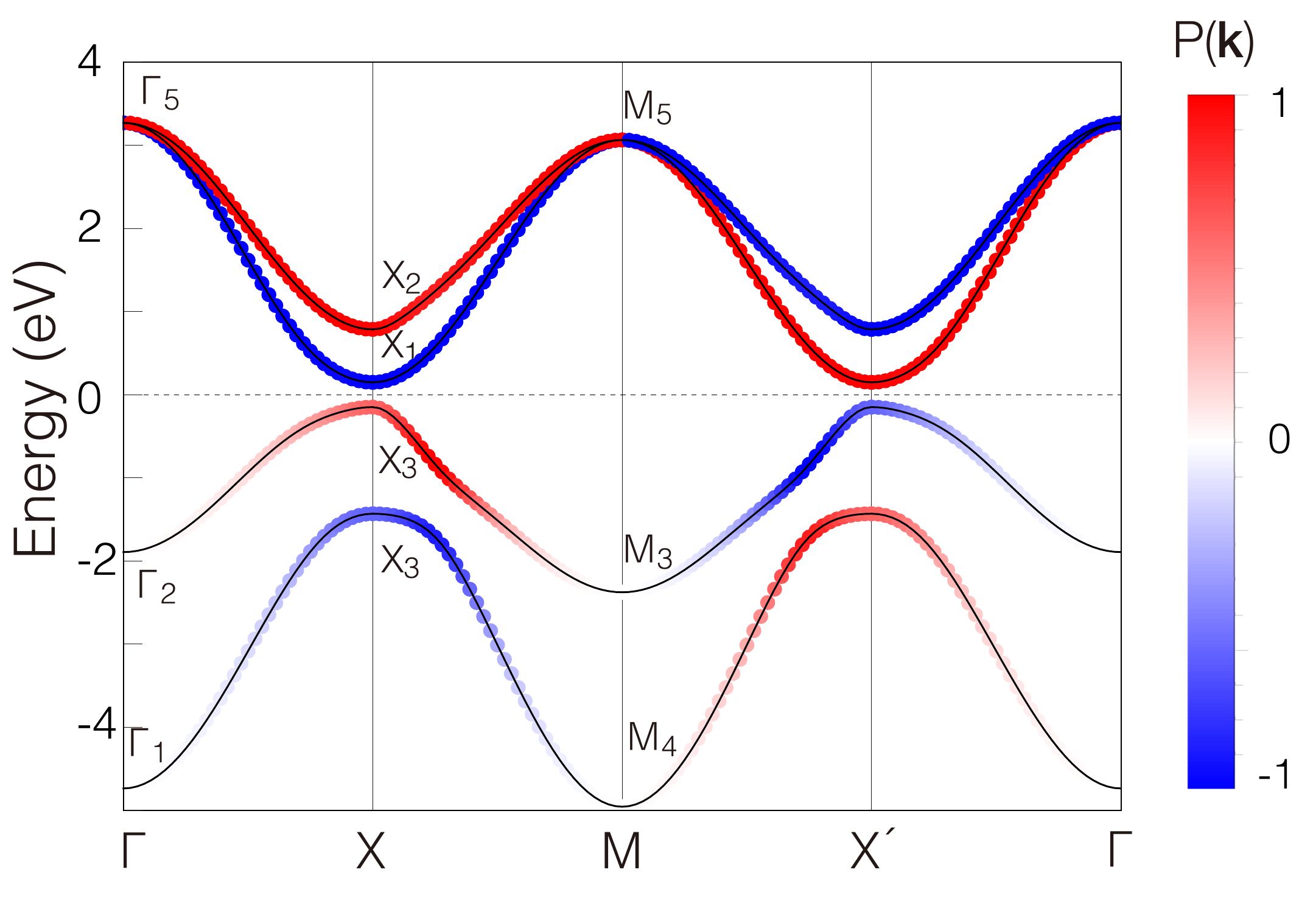}
	\caption{Layer polarization $P({\bm k})$ of TB  bands obtained from Hamiltonian (\ref{eq:ham}). The representation  of  the valence bands at high-symmetry points are labeled. }
	\label{Fig.3}
\end{figure}

\begin{figure}
	\includegraphics[width=8.8cm]{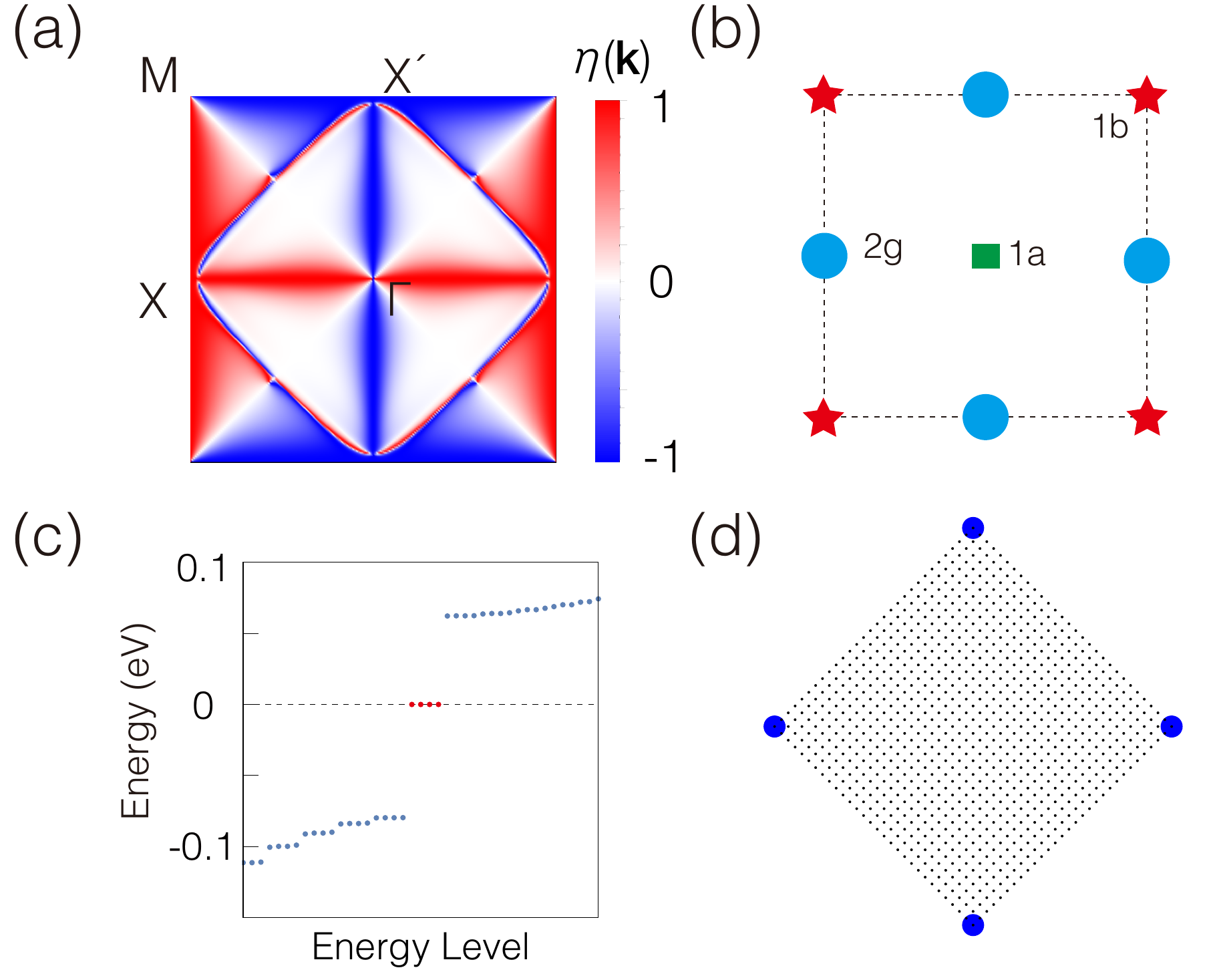}
	\caption{ (a)  $\eta({\bm k})$ of the TB model   (\ref{eq:ham}) in BZ. 
(b) Three  Wyckoff positions of the  monolayer $X_{2}Y$CO$_{2}$ in real space. In the TB model (\ref{eq:ham}), the $2g$ position is occupied by atoms, while $1a$ and $1b$ positions are empty. (c) Spectra of the  $X_{2}Y$CO$_{2}$ nanodisk, where four corner states (red dots) appear in the band gap. (d) The  distribution of the corner states in real space.}
	\label{Fig.4}
\end{figure}

Our four-band TB model can also reproduces the topological properties of the monolayer $X_{2}Y$CO$_{2}$, which is predicted  as a second-order topological insulator \cite{han2023cornertronics}. The topological properties of our  TB model can be directly diagnosed using the theory of topological quantum chemistry (TQC) \cite{TopologicalQuantumChemistry,TQC2017PRX,TOC2017NC,TOC2020Sci}.
As shown in Fig. \ref{Fig.3}, the irreducible representations of valence band at all the high-symmetry points are calculated as $\Gamma_{1}\oplus\Gamma_{2}+X_{3}\oplus X_{3}+M_{3}\oplus M_{4}$, which are induced by the $d_{x^{2}-y^{2}}$ and $d_{z^{2}}$ orbital located at $1b$ Wyckoff position [see Fig. \ref{Fig.4}(b)]. However, in the four-band TB model, the $1b$ Wyckoff position is empty. This indicates that the four-band TB model established  here  must be nontrivial.
According to the classification of higher-order topology \cite{TopologicalQuantumChemistry,highorderSci}, it is  a second-order topological insulator (SOTI).

A characteristic of the SOTI is the presence of corner states at specific corners of SOTI nanodisk. Here, based on the TB model, we calculate the spectrum for a nanodisk with $15\times15$ unit cells whose edges are on $110$ and $1\overline{1}0$ directions. The results are plotted in Fig. \ref{Fig.4}(c-d), where  four degenerate corner states in the gap of bulk state can be clearly observed. The
degeneracy of these four corner states is protected by the $S_{4z}$
symmetry of system.
Similar to the low-energy bands in bulk, the four corner states also have strong layer polarization.
Again, due to the $S_{4z}$ symmetry, the layer polarization of the corner states at $x$ and $y$ axis are opposite, consistent with the previous work\cite{han2023cornertronics}.

\section{Phase transitions \label{sec:IV}}

In addition to reproducing the low-energy physics of monolayer
$X_{2}Y$CO$_{2}$, the four-band TB model is itself physically interesting, and can host many topological phases under external fields. One can expect that these topological phases may be realized in the monolayer $X_{2}Y$CO$_{2}$ under suitable conditions.

\begin{figure}
\includegraphics[width=8.8cm]{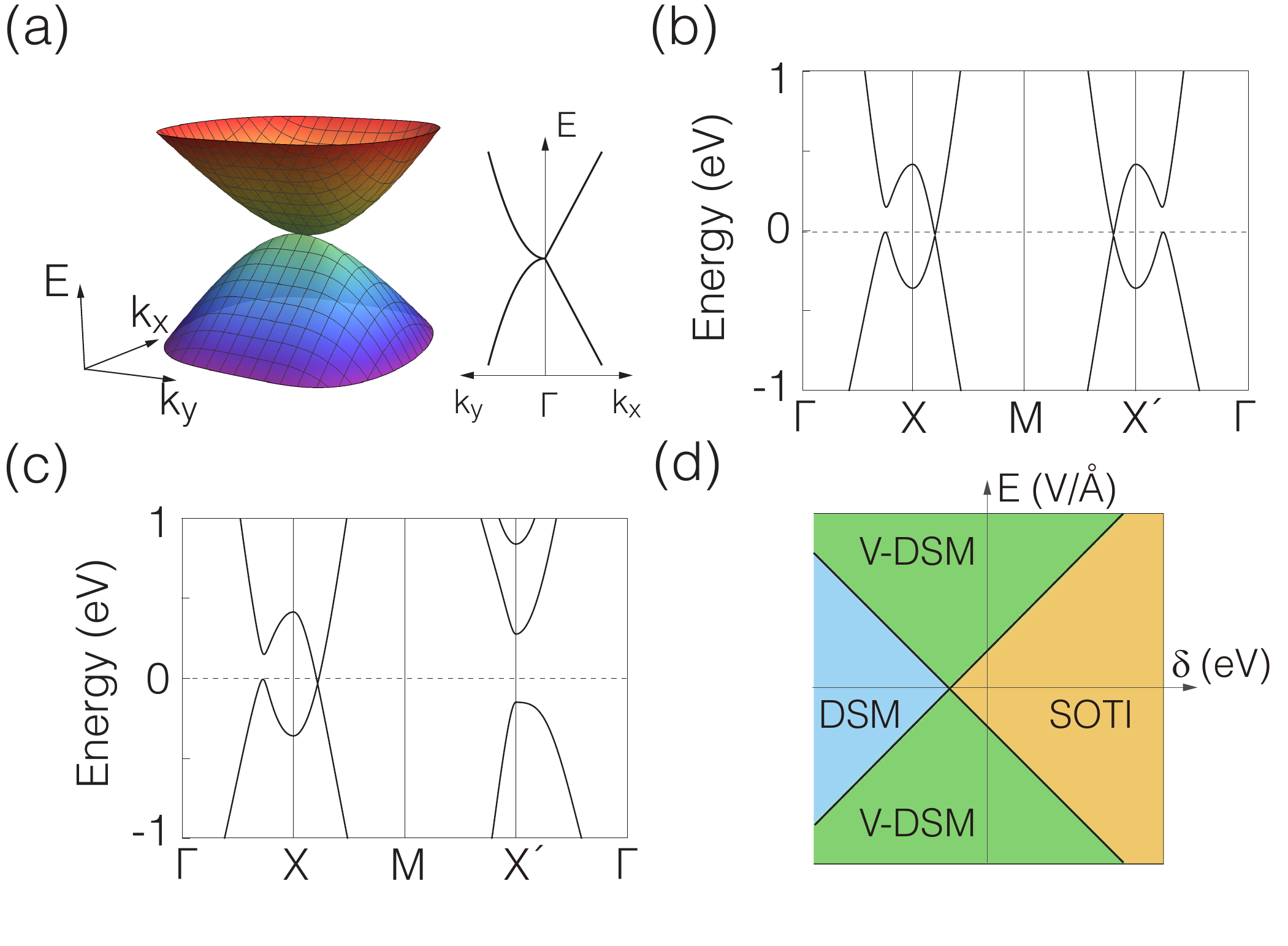}
\caption{(a) On the left, a 2D band structure illustrating a semi-Dirac point, and on the right, the dispersion of the semi-Dirac point's energy bands along high-symmetry paths. (b) and (c) are typical bands of $X_{2}Y$CO$_{2}$ in DSM and VPTDSM phase. (d) phase diagram of $X_{2}Y$CO$_{2}$ under gate electric field and symmetry-preserving perturbation. (b)(c)(d) are obtained using TB model of TiSiCO}
\label{Fig.5}
\end{figure}

Owing to the strong VLC effect, the most convenient way to control the bands of the TB model is applying a gate electric field normal to the plane of system, as it can produce an opposite electrostatic potential for the top and bottom atoms. Approximately, the effect of gate
electric field can be incorporated in the TB model (\ref{eq:ham}) by introducing  the an on-site energy term,
\begin{equation}\label{alpha}
{\cal{H}}_{E}=\alpha E\left(\begin{array}{cccc}
1& 0 & 0 & 0\\
0 & 1 & 0 & 0\\
0 & 0 & -1 & 0\\
0 & 0 & 0 & -1
\end{array}\right),
\end{equation}
where $E$ is the electric field and $\alpha$ is a real parameter depending on the material details, like  the separation of the top and bottom $X$ atoms, layer polarization of valley states, and  the screening effect. The values of $\alpha$ for different monolayer $X_{2}Y$CO$_{2}$ are listed  in Table \ref{tab.3}, which is extracted from the first-principle calculations [see Appendix \ref{App2}]. When $E$ is finite, both  $S_{4z}$  and $C_{2,110}$ symmetries of the system are broken, rendering the two valleys $X$ and $X'$ non-equivalent. As $E$ increases (assuming $E>0$), the band gap in $X$ valley decreases while that in $X'$ valley becomes lager. At a critical value $E=E_{c}$, the conduction and valence bands touch at $X$ valley, forming a semi-Driac point \cite{pardo2009half,MooreSemiDirac,semi-DiracLLs}.
Interestingly, the semi-Driac point exhibites a linear dispersion along $k_x$ direction but  a quadratic dispersion along $k_y$ direction [see Fig. \ref{Fig.5}(a)], and can be considered as a critical point where two conventional Dirac points merge together.
With the continuous increase of the gate field, the semi-Driac point splits  into two conventional Dirac points located at $M$-$X$ path, as illustrated in Fig. \ref{Fig.5}(c). Since both Dirac points reside around $X$ valley, we term this  phase as valley-polarized topological Dirac semimetal (V-DSM).

{\global\long\def\arraystretch{1.4}%
	\begin{table}
		\caption{Fitted parameter $\alpha$ [in Eq. (\ref{alpha})]. $\alpha$ is in the unit of \AA. }
		\begin{ruledtabular} %
			\begin{tabular}{lllllllll}
				& TiSiCO & TiGeCO & ZrSiCO & ZrGeCO & HfSiCO & HfGeCO\tabularnewline
				\hline
				$\alpha$ &-0.204  &-0.182  &-0.237  &-0.241  &-0.267  &-0.249 \tabularnewline
		\end{tabular}\end{ruledtabular}
		\label{tab.3}
	\end{table}
}

The four-band TB model  can also be tuned by symmetry-preserving perturbations like  biaxial strain, which changes the value of the parameters in the original Hamiltonian (\ref{eq:ham}).
Consider a symmetry-preserving  perturbation
\begin{equation}	{\cal{H}}_{\delta}=\delta(\mathrm{cos}k_{x}-\mathrm{cos}k_{y})\left(\begin{array}{cccc}
		1 & 0 & 0 & 0\\
		0 & 1 & 0 & 0\\
		0 & 0 & -1 & 0\\
		0 & 0 & 0 & -1
	\end{array}\right),
\end{equation}
corresponding to the situation that  the  parameter $t_{i,\alpha}$ ($i={1,2}$ and $\alpha={x,y}$) by $\delta$ has been changed, while  the other parameters unchange.

This perturbation changes the band gap at both valleys. Particularly,  the band gap of system decreases when $\delta<0$, and closes at a  critical value $\delta=\delta_{c}$, in such case, there exist two semi-Driac points residing at $X$ and $X'$ valleys, respectively.
When  $\delta<\delta_{c}$, the two semi-Driac points become four symmetry-connected conventional Dirac points, and the system becomes a topological Dirac semimetal (DSM), as shown in Fig. \ref{Fig.4}(c). The phase diagram of the TB model under these two perturbations $E$ and $\delta$ is plotted in Fig. \ref{Fig.4}(d).

\section{Conclusion \label{sec:V}}
In this work, we construct a four-band TB model for the TiSiCO-family monolayers ($X_{2}Y$CO$_{2}$) based on the $d$-orbitals of the two $X$ atoms.
Via the  theory of band representation, we show this four-band  model is the minimal one that can capture the low-energy physics of the TiSiCO-family monolayers.
Our TB model includes both  inter-layer and intra-layer NN hoppings, and the hopping parameters are fitted to first-principle calculations by gradient descent method.
Consequently, our model accurately reproduces the  energy dispersion and the layer polarization of the bands around   X and X$'$ valleys.
Our model can also describe the valley-contrasted linear dichroism of the  monolayer $X_{2}Y$CO$_{2}$.
Furthermore, we demonstrate that the TB model is a SOTI, and exhibits  topological  corner states in its nanodisk.
These results are consist with that calculated from first-principle calculations.

We then investigate the possible phase transitions of the TB model under different perturbations. Under a  gate field and  biaxial strain, the TB model is transformed from a SOTI to two distinct phases: valley-polarized topological Dirac semimetal and conventional topological Dirac semimetal.
Therefore, our TB model  not only effectively describes the low-energy properties of monolayer $X_{2}Y$CO$_{2}$, which greatly simplify the further study on monolayer $X_{2}Y$CO$_{2}$ materials, but  also can be used to study the interplay between valley physics and higher-order topology.

\acknowledgements
The authors thank J. Xun for helpful discussions.
This work was supported by the National Key R\&D Program of China (Grant No. 2020YFA0308800), the NSF of China (Grants Nos. 12004035, 12061131002 and 12234003), and the  National Natural Science Fund for Excellent Young Scientists Fund Program (Overseas).

\appendix
\section{Band analysis of the TiSiCO-family monolayers }\label{App1}

\begin{figure*}
	\includegraphics[width=14cm]{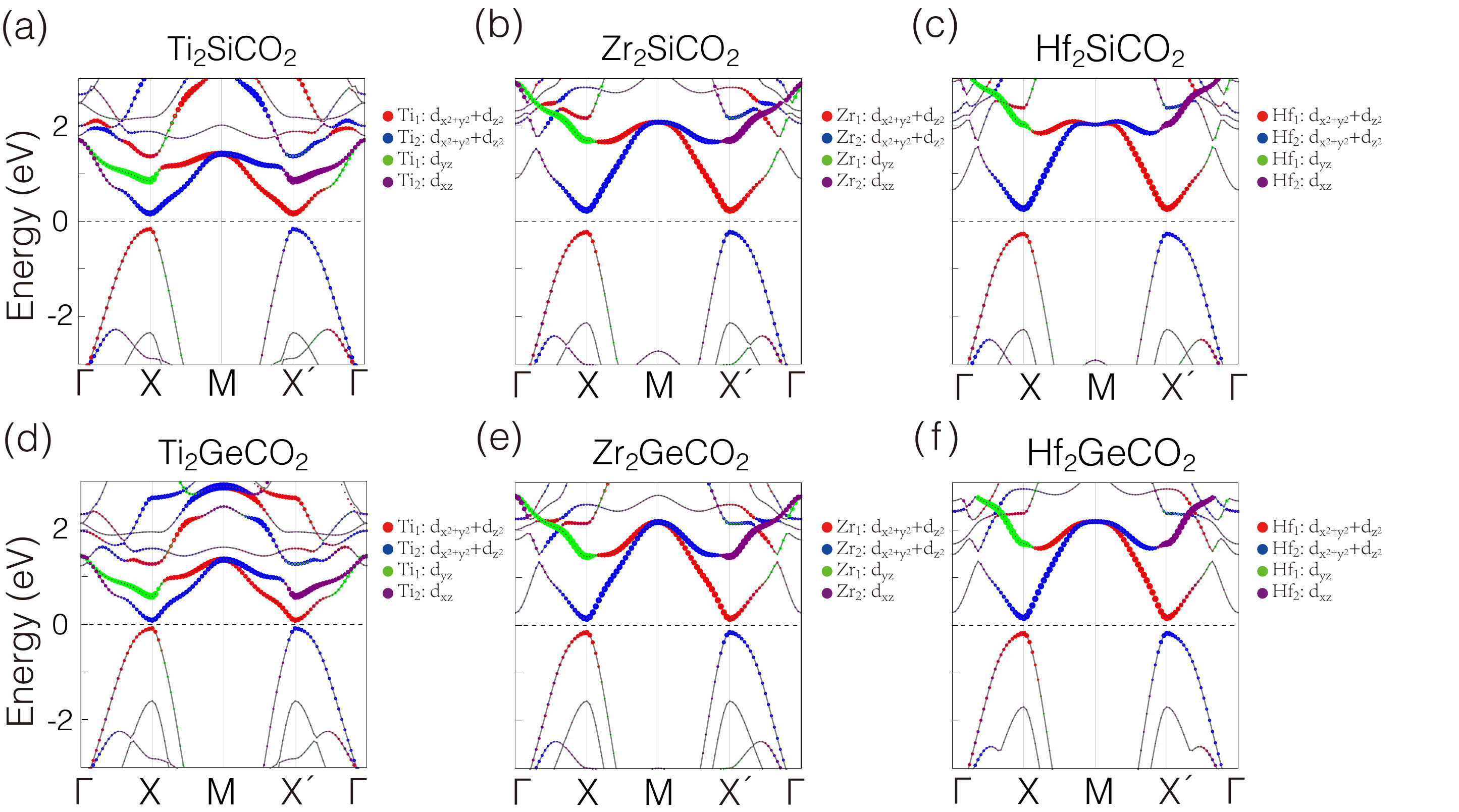}
	\caption{(a-f) Electronic band structure of monolayer $X_{2}Y$CO$_{2}$  ($X=$ Ti, Zr, Hf; $Y=$ Si, Ge) without SOC. The size of the colored circles is proportional to the weight of projection onto atomic orbitals.}\label{Fig.S1}
\end{figure*}

Figure \ref{Fig.S1} shows the electronic band structures of the  monolayers $X_{2}Y$CO$_{2}$  without SOC. The  projection of the electronic bands
onto atomic orbitals is  also presented.
 One observes that they have  similar features as those discussed in the main text for monolayer Ti$_2$SiCO$_2$.

\section{Gate field control of valley states in  TiSiCO-family monolayers} \label{App2}

\begin{figure*}
	\includegraphics[width=14cm]{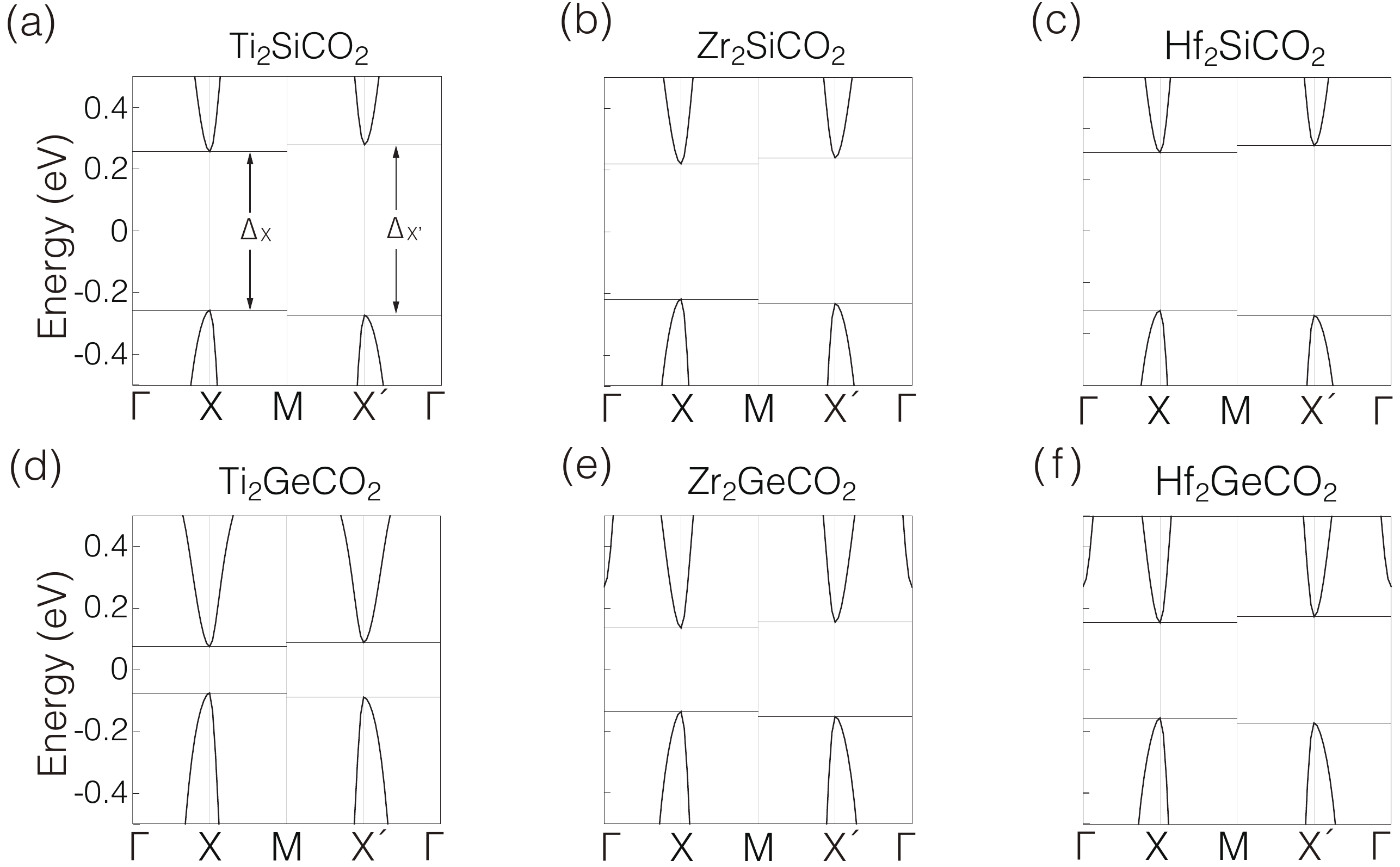}
	\caption{(a-f) Electronic band structure of monolayer $X_{2}Y$CO$_{2}$  ($X=$ Ti, Zr, Hf; $Y=$ Si, Ge) under a  gate field of $0.05$ eV/\AA~,obtained from first-principle calculations. The gap at $X$ and $X'$ point is denoted as $\Delta_{X}$ and $\Delta_{X'}$. $\alpha$ is obtained by comparing the values of  $\Delta_{X}-\Delta_{X'}$ from  TB model (\ref{eq:ham}) with Eq. (\ref{alpha}) and that from first-principle calculations. }\label{Fig.S2}
\end{figure*}

Due to VLC, a gate-field control of the  valley  polarization can be realized in  the  monolayer $X_{2}Y$CO$_{2}$.
The electronic band of the  monolayer $X_{2}Y$CO$_{2}$ under a gate field of $0.05$ eV/\AA ~ are shown in Fig. \ref{Fig.S2}, from which the coefficient $\alpha$ in Eq. (\ref{alpha}) can be obtained.

\bibliography{ref}
\end{document}